\documentclass[amsthm]{my_autart_long}
\usepackage{graphicx}          % Include this line if your%%
\usepackage{subfigure}
\usepackage{multirow,float}
\usepackage[cmex10]{amsmath,mathtools}
\usepackage{amsfonts,amssymb}  % assumes amsmath package installed
\usepackage{times}
\usepackage[usenames]{color}
\usepackage[colorinlistoftodos]{todonotes}
\usepackage{algorithm}
\usepackage{algorithmic}

\usepackage[numbers,sort&compress]{natbib}
\usepackage{ifpdf}
\usepackage{ulem}
\usepackage{appendix}
\usepackage{enumitem}

\newcommand{\be}{\begin{equation}}
\newcommand{\ee}{\end{equation}}
\newcommand{\bee}{\begin{enumerate}}
\newcommand{\eee}{\end{enumerate}}
\newcommand{\ldet}{\mathrm{ldet}}
\newcommand{\Cov}{\mathrm{Cov}}

\begin{document}
\allowdisplaybreaks

\begin{frontmatter}
\title{Mutual Information-Based Planning for Informative Windowed Forecasting of Continuous-Time Linear Systems}

\thanks[footnoteinfo]{Corresponding author; Tel: +82-42-350-3727; Fax: +82-42-350-3710; E-mail: hanlimc@kaist.ac.kr}

\author[kaist]{Han-Lim Choi$^\star$}%\ead{hanlimc@kaist.ac.kr} %and    % Add the
%\author[mit]{Jonathan P. How}\ead{jhow@mit.edu}               % e-mail address

\address[kaist]{291 Daehak-ro, Yuseong, Deajeon 305-701, Republic of Korea.}
%\address[mit]{77 Massachusetts Ave., Cambridge, MA 02139, USA.}  % Please supply


%XX shold title be ``Uncertainties'' or ``Uncertainty'' ? the latter
%seems more appropriate to me {\CF agreed} XX

%%%%%%%%%%%%%%%%%%%%%%%%%%%%%%%%%%%%%%%%%%%%%%%%%%%%%%%%%%%%%%%%%%%%%%%%%%%%%%%%
\begin{abstract}
This paper presents expression of mutual information that defines the information gain in planning of sensing resources, when the goal is to reduce the forecast uncertainty of some quantities of interest and the system dynamics is described as a continuous-time linear system. The method extends the smoother approach in \cite{ChoiHow_Auto10} to handle more general notion of verification entity - continuous sequence of variables over some finite time window in the future. The expression of mutual information for this windowed forecasting case is derived and quantified, taking advantage of underlying conditional independence structure and utilizing the fixed-interval smoothing formula with correlated noises. Two numerical examples on (a) simplified weather forecasting with moving verification paths, and (b) sensor network scheduling for tracking of multiple moving targets are considered for validation of the proposed approach.
\end{abstract}
\end{frontmatter}

%\input{HPC_IntroBrief}
%%%%%%%%%%%%%%%%%%%%%%%%%%%%%%%%%%%%%%%%%%%%%%%%%%%%%%%%%%%%%%%%%%%%%%%%%%%%%%%%
\section{Introduction}

\parskip=.1in

Planning on utilization of sensing resources to gather information out of environment has been spotlighted in many contexts, the objective of this planning often being uncertainty reduction of some entities of interest -- termed \textit{verification} entities herein. Mutual information has been one of the most popular metrics adopted to define/represent this objective for various context: tracking of kinematic variables of moving targets by measurement along mobile sensor trajectories~\cite{Grocholsky_PhD02, HoffmanTomlin_TAC10}, weather forecast improvement over some region of interest in the future with UAV sensor networks~\cite{ChoiHow_TCST11, ChoiHow_SJ11, ChoiHow_Auto10}, prediction accuracy in spatially distributed field described by Gaussian processes~\cite{KrauseGuestrin_JMLR08}, informative management of deployed fixed sensor networks~\cite{Williams_TSP07, ChoiHowBarton_OPTL13}, adaptive landmark selection in simultaneous localization and mapping of mobile robots~\cite{KretzschmarStachiniss_IJRR12}, and Bayesian belief propagation over the grid-based search space~\cite{Julian_IJRR12}.

%This popularity results in part from that mutual information is a more natural quantity of information (than entropy) that can be applied to general random entities (i.e., random variables, random processes, random functions, random systems), and in part from that often times sufficient statistics (rather than the actual data) can be used to compute the mutual information. In addition, recent progress on efficient Bayesian inference methods has contributed to effective quantification of mutual information in various contexts.

While many of these mutual information-based planning studies have dealt with the case where the verification time is same or just one time-step further of the planning horizon, there is a class of problem termed \textit{informative forecasting} that takes particular care for the case where the verification time is significantly greater than the mission horizon. Although less popular in the literature, the informative forecasting problem can handle applications such as (i) adaptive sampling in the context of numerical weather prediction considers design of sensor networks deployed in the near future (e.g., in 24 hours) while the goal is to improve forecast in the far future (e.g., 3-5 days later), and (ii) prediction of indoor contaminant distribution in some future time with wireless indoor sensor networks taken over short period of time.  The present author has presented methods to efficiently but correctly quantify the mutual information in this context of informative forecasting for discrete selection case \cite{ChoiHow_TCST11}, discrete constrained path design \cite{ChoiHow_SJ11}, and continuous trajectory planning\cite{ChoiHow_Auto10}, taking advantage of underlying properties of mutual information.

This paper extends the approach in \cite{ChoiHow_Auto10} in that a more general notion of verification quantities is introduced.  For some applications, it may make more sense to reduce uncertainty in the entities of interest over some finite window of time instead of a single particular time instance (for example, weather forecast over the weekend).  The smoother form in \cite{ChoiHow_Auto10} cannot directly be used for this \textit{windowed forecasting} case, because the mutual information between two continuous random processes (as opposed to one finite-dimensional random vector and one random process) needs to be calculated. This paper presents a formula for the mutual information for this windowed forecasting that is indeed quite similar to the form in \cite{ChoiHow_Auto10}, while the only difference is in the process of calculating the conditional initial covariance conditioned on the verification entity. An optimal-control based method for fixed-interval Kalman smoothing  with correlated noise in\cite{MehraBryson_TAC68} is adopted for this calculation. Two numerical examples are presented to validate the proposed method for the cases with (i) time-varying verification entity, and (ii) high-order differentiable verification entity. While a preliminary work~\cite{ChoiHow_ICARCV10} proposed a relevant concept of the windowed forecasting,  this article presents elaborated and corrected theoretical results for a more general problem setting and also provides much more sophisticated numerical case studies.

\section{Problem Description}
\subsection{Continuous-Time Linear System Model} \label{sec:model}
Consider the dynamics of objects/environment of interest with a finite
dimensional state vector $X_t \in \mathbb{R}^{n_X}$ that is
described by the following linear (time-varying) system:
\be %
\dot{X_t} = A(t)X_t + B(t) W_t \label{eq:lindyn}
\ee %
where $W_t \in \mathbb{R}^{n_W}$ is a zero-mean Gaussian process
noise with $\mathbb{E}[ W_t W_s' ] = \Sigma_W \delta(t-s),~~
\Sigma_W \succ 0$, which is independent of $X_t$. The prime sign
($'$) denotes the transpose of a matrix. The initial condition of
the state, $X_0$ is normally distributed as $X_0 \sim
\mathcal{N}(\mu_0, P_0),~ P_0 \succ 0$.

The system (\ref{eq:lindyn}) is observed by sensors with additive
Gaussian noise and admits the following measurement model for $Z_t
\in \mathbb{R}^{n_Z}$:
\be %
Z_t = C(t)X_t + N_t \label{eq:linobs}
\ee %
where $N_t \in \mathbb{R}^{n_Z} $ is zero-mean Gaussian with
$\mathbb{E}[N_t N_s' ] = \Sigma_N \delta(t-s),~~\Sigma_N \succ 0$,
which is independent of $X_t$ and $W_s,~\forall s$. Also, a
measurement history over the time window $[t_1,t_2]$ is defined as
\be %
\mathcal{Z}_{[t_1,t_2]} = \{ Z_{t} : t \in [t_1,t_2] \}.
\ee %

\begin{defn}
The verification variables are a possibly time-varying linear combination of the state variables whose uncertainty reduction is of interest:
\be %
V_t = M_V(t) X_t \in \mathbb{R}^{n_V} \label{eq:Vt}
\ee %
with $M_V(t) \in \mathbb{R}^{n_V \times n_X} $ termed as \textit{verification matrix}, which is assumed to be \textit{differentiable} herein. A continuous sequence of (time-varying) verification variables, termed \textit{verification path},  is also defined as:
\be %
\mathcal{V}_{[t_1, t_2]} = \{ V_t : t \in [t_1, t_2] \}.
\ee %
%\qed
\end{defn}

%\begin{assum}
%This work assumes that the verification matrix $M_V(t)$ is differentiable in time. %\qed
%\end{assum}

\subsection{Informative Pointwise Forecasting}
One case of interest is when  the verification entity is the verification {\it variables} at some fixed verification time, $T$. In this case, the informative forecasting problem can be written as the following optimization:
\be %
\max_{\mathcal{Z}_{[0,\tau]}}~\mathcal{I}(V_T;
\mathcal{Z}_{[0,\tau]}) \tag{\textbf{IPF}} \label{eq:ipf}
\ee %
with some $\tau \in [0,T]$, where $\mathcal{I}(Y_1;Y_2)$ denoted the mutual information
between two random quantities  $Y_1$ and $Y_2$ (e.g. random
variables, random processes, random functions), which represents entropy reduction of $Y_1$ by knowledge of$Y_2$ (or equivalently, entropy reduction of $Y_2$ by $Y_1$~\cite{inftheory}. Thus, (\ref{eq:ipf}) finds the (continuous) measurement sequence
over $[0,\tau]$ that is expected to result in largest reduction of entropy in $V_T$.
%The present author~\cite{ChoiHow_Auto10} developed methods to efficiently and correctly quantify the mutual information for this first type of informative forecasting.
% An example of this type of planning is to design sensing paths for UAV sensor platforms operating in the near future (e.g., within 24 hours) to improve weather forecast over some verification region at a given time of interest in the future (e.g., in 5 days).

Exploiting conditional independence, we proposed an expression for the mutual information, $\mathcal{I}(V_T ; \mathcal{Z}_{[0,\tau]})$,  as the difference between the unconditioned and the conditioned mutual information for a filtering problem~\cite{ChoiHow_Auto10}:
\be %
 \mathcal{I}(V_T ; \mathcal{Z}_{[0,\tau]} ) = \mathcal{I}
(X_{\tau};\mathcal{Z}_{[0,\tau]}) - \mathcal{I}(X_{\tau};
\mathcal{Z}_{[0,\tau]} | V_T ). \label{eq:info_equiv}
\ee %
With (\ref{eq:info_equiv}), the \textit{smoother form} of the mutual
information for forecasting is derived as
\be %
\begin{split}
\mathcal{I}(V_T;\mathcal{Z}_{[0,\tau]}) &= \mathcal{I}(X_{\tau}
;\mathcal{Z}_{[0,\tau]}) - \mathcal{I}(X_{\tau} ;
\mathcal{Z}_{[0,\tau]}
|V_T) \\
& = \mathcal{J}_0(\tau) - \textstyle{\frac{1}{2}} \, \ldet
(I+Q_X(\tau) \Delta_S(\tau)) \label{eq:smform2}
\end{split}
\ee %
with $\mathcal{J}_0 \triangleq \textstyle{\frac{1}{2}} \ldet
S_{X|V} - \textstyle{\frac{1}{2}} \ldet S_{X}$ and $\Delta_S
\triangleq S_{X|V}- S_{X}$, where $\ldet$ stands for $\log \det $ of a positive definite matrix. The matrices $S_X(\tau) \triangleq \Cov^{-1}( X_\tau )$, $S_{X|V}(\tau)\triangleq \Cov^{-1}( X_\tau |V_T)$, and $Q_X(\tau) \triangleq \Cov( X_\tau |\mathcal{Z}_{[0,\tau]})$ are determined by the following matrix differential equations:
\begin{align}
&\dot{S}_{X} = - S_{X} A - A'S_{X} - S_{X} B \Sigma_W B' S_{X} \label{eq:sx}\\
&\dot{S}_{X|V} = S_{X|V} B \Sigma_W B' S_{X|V} - S_{X|V} (A + B \Sigma_W B' S_X) \notag \\
&\qquad \qquad  - (A + B \Sigma_W B' S_X)'S_{X|V}   \label{eq:sxv}\\
& \dot{Q}_X = A Q_X + Q_X A' + B \Sigma_W B' - Q_X C'\Sigma_N^{-1} C
 Q_X   \label{eq:qx}
\end{align}
with initial conditions $S_{X}(0) = P_0^{-1},~S_{X|V}(0) =
P_{0|V}^{-1}$, and $Q_X(0) = P_0$. The conditional initial covariance $P_{0|V} \triangleq \Cov(X_0
|V_T) \succ 0$ can be calculated in advance by a fixed-point
smoothing process, or simply by
$$
P_{0|V}  = P_0  - P_0 \Phi_{(T,0)}' M_V' \left[ M_V P_X(T) M_V'
\right]^{-1}  M_V \Phi_{(T,0)} P_0
$$
where $\Phi_{(t_2,t_1)}$ is the state transition matrix from $t_1$ to $t_2$, which becomes
$e^{A(t_2 - t_1)}$ for the time-invariance case.

In \cite{ChoiHow_Auto10}, we demonstrated that the
smoother form is preferred to the filter form, which explicitly calculates the prior and the posterior entropies of $V_T$ by integrating the Lypunov and the Riccati equation over $[0,T]$,  in terms of the computational efficiency and accessibility to on-the-fly knowledge of information accumulation.

\iffalse
 It was demonstrated in \cite{ChoiHow_Auto10} that the
smoother form is preferred to the filter form  in terms of the
computational efficiency and accessibility of the on-the-fly
knowledge of the accumulated information. For comparison, the filter form is given by:
\be %
\begin{split}
&\mathcal{I}(V_T; \mathcal{Z}_{[0,\tau]})  \\&~~ = \textstyle{\frac{1}{2}} \ldet ~(M_V P_X(T) M_V')-
\textstyle{\frac{1}{2}} \ldet~  (M_V Q_X(T) M_V')
\end{split}
\ee %
where $P_X(T) \triangleq \Cov(X_T)$ and $Q_X(T) \triangleq \Cov (X_T|\mathcal{Z}_{[0,\tau]})$ are obtained by
\begin{align}
 \dot{P}_X(t) &= A(t)P_X(t) + P_X(t) A'(t) + \Sigma_W \notag \\
 %\label{eq:pxT}
 \dot{Q}_X &= A Q_X + Q_X A'
 + \Sigma_W - \mathbb{I}_{[0,\tau]} Q_X C'\Sigma_N^{-1} C
 Q_X \notag %\label{eq:qxT}
\end{align} %
with initial conditions $P_X(0) = Q_X(0) = P_0$.  $\mathbb{I}_{[0,\tau]}(t): \mathbb{R}_{+} \mapsto \{0,1\}$
is the indicator function that is unity for $t \in [0,\tau]$ and
zero elsewhere.  The filter form explicitly represents the mutual information as the difference between the prior and the posterior entropy of the verification variables; these entropy terms are propagated from the conditional entropy of the state at $\tau$. Note that for the filter form the Riccati equation (of $Q_X$) needs to be integrated over $[0,T]$, while it is integrated only over $[0,\tau]$ for the smoother form; this yields computational saving of a factor of $T/\tau$ of the use of the smoother form for informative planning problems.
\fi

\subsection{Informative Windowed Forecasting}

\iffalse
The planning problem for informative forecasting addresses design of
sensing paths for mobile sensors (or equivalently scheduling sequence of distributed sensors) over some given time window $[0,\tau]$ in the near future to reduce the uncertainty in some verification entities in the far future.  One popular notion to quantify the expected amount of uncertainty reduction of the verification entities by the measurements taken along the sensing paths  is the mutual information (between the verification entities and the measurement sequence).
\fi

This paper newly considers  a more general version of the informative forecasting problem
where the entity of interest is the verification path over some time
window $[T_i, T_f]$.  This generalized problem can
be written as:
\begin{align*}
&\max_{\mathcal{Z}_{[0,\tau]}}~\mathcal{I}(\mathcal{V}_{[T_i,T_f]};
\mathcal{Z}_{[0,\tau]})  \tag{\textbf{IWF}} \label{eq:iwf}
\end{align*}
with $V_t = M_V(t) X_t,~~ t \in [T_i, T_f]$. This formulation allows for handling more diverse types of sensing missions such as to  weather forecast over the coming weekend, better predicting behaviors of some targets of interest between 9am and noon, and so on. Note also that because the verification matrix is allowed to be time-varying, forecasting along a given path can be dealt with (e.g., weather forecasting along my itinerary in the weekend).

\iffalse
In addition, the formulation in (\ref{eq:iwf}) can also be used to address more  robust decision making for (\ref{eq:iwf}), specifically with setting $T_f = T_i + \epsilon$ with some small $\epsilon$.  For sensor networks operating in dynamic environments, the mission specifications might be unclear and there might be some uncertainty in when the verification time should be; then, (\ref{eq:iwf}) can be posed to account for such uncertainty in the verification time.
\fi

However, the generalization (\ref{eq:iwf}) gives rise to challenges in quantifying the mutual information in the objective function.  For discrete-time representation, in which $\mathcal{Z}_{[0,\tau]}$ and $\mathcal{V}_{[T_i, T_f]}$ can be represented by finite dimensional random vectors, the similar generalization as in (\ref{eq:iwf}) would not incur any additional difficulty in computation of mutual information other than computational cost due to the increased dimension of the verification entity. Thus, quantification and optimization methods developed for the discrete-time counterpart of (\ref{eq:ipf}) can trivially extended for the discrete-time counterpart of (\ref{eq:iwf}).

In contrast, for continuous-time representation considered herein, the objective term in (\ref{eq:iwf}) is mutual information between two continuous-time random processes, while that in (\ref{eq:ipf}) is mutual information between a finite-dimensional random vector and a continuous-time random process. If the mutual information is computed as a difference between the prior and the posterior entropy of one of the two random processes, a mechanism to calculate an entropy of a continuous random process is needed. Although there have been researches on the calculation of entropy of a continuous random process~\cite{Termonia_PRA84, Faure_PhysicaD98}, these approaches were to statistically estimate the entropy value from experimentally-obtained time series, and thus are not suitable for quantifying the information by a  future measurement that is not taken yet at the decision time.

\section{Main Results} \label{sec:mutual_info}

\subsection{Mutual Information for Windowed Forecasting} \label{sec:main_iwf}
\iffalse
First note that the mutual information for (\ref{eq:iwf}) cannot be readily quantified in the filtering framework that computes
the difference between the prior and posterior entropies of the entities of interest. In (\ref{eq:iwf}), the verification entity is a continuous-time random process $\mathcal{V}_{[T_i,T_f]}$, and there is no direct way to calculate the entropy of some Gaussian random process over an arbitrary time window without the actual time series of that signal. For the informative forecasting problem in this work, such time series are not available at the decision time, and the decision should be made based on the statistics of the random quantities rather than their actual values.
\fi

As the present author presented in \cite{ChoiHow_Auto10}, the mutual information in (\ref{eq:ipf}), $\mathcal{I}(V_T; \mathcal{Z}_{[0,\tau]})$, can be expressed in terms of unconditioned and conditional entropies of finite-dimensional random vectors, exploiting the popular conditional independence between the future and the past given the present. For linear Gaussian cases, these entropy terms  are represented by functions of covariance matrices. Inspired by this observation, this paper provides an expression for the mutual information in (\ref{eq:iwf}) as a function of covariance matrices for some finite-dimensional random vectors as follows.

\begin{prop}
If unconditioned and conditioned covariances of the initial state vector,  $P_0 \triangleq \Cov (X_0) $ and $P_{0|\mathcal{V}} \triangleq  \Cov (X_0 | \mathcal{V}_{[T_i, T_f]})$, are available. The mutual information $\mathcal{I}(\mathcal{V}_{[T_i, T_f]} | \mathcal{Z}_{[0,\tau]})$ can be obtained as:

\begin{equation}
\begin{split}
\mathcal{I}(\mathcal{V}_{[T_i, T_f]};\mathcal{Z}_{[0,\tau]}) &= \mathcal{I}(X_{\tau}
;\mathcal{Z}_{[0,\tau]}) - \mathcal{I}(X_{\tau} ;
\mathcal{Z}_{[0,\tau]}
|\mathcal{V}_{[T_i, T_f]}) \\
& = \mathcal{J}_0^w(\tau) - \textstyle{\frac{1}{2}} \, \ldet
(I+Q_X(\tau) \Delta_S^w(\tau))  \label{eq:iwf_smoother}
\end{split}
\end{equation} %
with $\mathcal{J}_0 \triangleq \textstyle{\frac{1}{2}} \ldet
S_{X|\mathcal{V}} - \textstyle{\frac{1}{2}} \ldet S_{X}$ and $\Delta_S^w
\triangleq S_{X|\mathcal{V}}- S_{X}$. The matrices $S_X(\tau) \triangleq \Cov^{-1}( X_\tau )$, $S_{X|\mathcal{V}}(\tau)\triangleq \Cov^{-1}( X_\tau |\mathcal{V}_{[T_i, T_f]})$, and $Q_X(\tau) \triangleq \Cov( X_\tau |\mathcal{Z}_{[0,\tau]})$ are determined by integrating the following
matrix differential equations from time 0 to $\tau$:
\begin{align}
& \dot{S}_{X} = - S_{X} A - A'S_{X} - S_{X} B \Sigma_W B' S_{X} \label{eq:sx_w} \\
& \dot{S}_{X|\mathcal{V}} = S_{X|\mathcal{V}} B \Sigma_W B' S_{X|\mathcal{V}} - S_{X|\mathcal{V}} (A + B \Sigma_W B' S_X)\notag \\ & \qquad \qquad  - (A + B \Sigma_W B' S_X)'S_{X|\mathcal{V}}    \label{eq:sxv_w}\\
 & \dot{Q}_X = A Q_X + Q_X A' + B \Sigma_WB'  - Q_X C'\Sigma_N^{-1} C
 Q_X.   \label{eq:qx_w}
\end{align}

\begin{proof}
The similar conditional independence exploited to derive the smoother form for (\ref{eq:ipf}) can also be used for (\ref{eq:iwf}),
because the verification variables of a future time window $[T_i, T_f]$ is conditionally independent of the (past) measurement sequence
$\mathcal{Z}_{[0,\tau]}$, conditioned on the (current) state variables $X_{\tau}$. Thus, we have:
$$
\mathcal{I}(\mathcal{V}_{[T_i,T_f]};\mathcal{Z}_{[0,\tau]}) = \mathcal{I}(X_{\tau}; \mathcal{Z}_{[0,\tau]}) - \mathcal{I}(X_{\tau}; \mathcal{Z}_{[0,\tau]}|\mathcal{V}_{[T_i,T_f]}),
$$
which is derived using the fact that $\mathcal{I}(\mathcal{V}_{[T_i, T_f]}; \mathcal{Z}_{[0,\tau]}|X_{\tau}) = 0$ due to the conditional independence.  Notice that the first term in the left-hand side is identical to that in the expression for (\ref{eq:ipf}). The second term represents the difference between two conditional entropies, $\mathcal{H}(X_{\tau} | \mathcal{V}_{[T_i, T_f]})$ and $\mathcal{H}(X_{\tau} | \mathcal{V}_{[T_i,T_f]}, \mathcal{Z}_{[0,\tau]})$. Since the conditional distribution of a Gaussian vector conditioned on some Gaussian random process is still Gaussian, these two entropy expressions can be represented by $\log \det $ of the corresponding covariance matrices:
$$
\mathcal{I}(X_{\tau}; \mathcal{Z}_{[0,\tau]}|\mathcal{V}_{[T_i,T_f]}) = \textstyle{\frac{1}{2}} ( \ldet P_{X|\mathcal{V}} (\tau)
- \ldet Q_{X|\mathcal{V}}(\tau) )
$$
where $P_{X|\mathcal{V}}(\tau) \triangleq \Cov (X_{\tau} | \mathcal{V}_{[T_i, T_f]})$ and $Q_{X|\mathcal{V}}(\tau) \triangleq \Cov (X_{\tau}|\mathcal{V}_{[T_i,T_f]}, \mathcal{Z}_{[0,\tau]} )$.  Note that the smoother form for (\ref{eq:ipf}) utilized the (symmetric) two-filter approach to fixed-interval smoothing in \cite{smoothing_willsky} to express the conditional covariance $Q_{X|V}(\tau) \triangleq \Cov (X_{\tau} | V_T)$ in terms of $P_X(\tau)$, $Q_X(\tau)$, and $P_{X|V}(\tau)$. The key insight \cite{smoothing_willsky} has identified is: the information matrix for the fixed interval smoothing, i.e., $Q_{X|V}^{-1}$, consists of the information from the past measurement, i.e., $Q_{X}^{-1}$, and the information from the future (fictitious) measurement, i.e., $P^{-1}_{X|V}$ in this context, minus the double-counted information from the underlying dynamics, i.e., $P^{-1}_X$, thus yielding $
Q_{X|V}^{-1}(\tau) = Q_X^{-1}(\tau) + P_{X|V}^{-1}(\tau) - P_X^{-1}(\tau).
$ Notice that this key insight in fixed-interval smoothing still holds even in case the future measurement (of $V_t$)  is taken over a finite window $[T_i,T_f]$. Thus, we have
$$
Q_{X|\mathcal{V}}^{-1}(\tau) = Q_X^{-1}(\tau) + P_{X|\mathcal{V}}^{-1}(\tau) - P_X^{-1}(\tau),
$$
and thus the only term that did not appear in (\ref{eq:ipf}) is  $P_{X|\mathcal{V}}$
(or equivalently, $S_{X|\mathcal{V}} \triangleq P_{X|\mathcal{V}}^{-1}$).  
This quantity this conditional covariance (or inverse covariance) can be obtained by integrating a Lyapunov-like equation in the same form as (\ref{eq:sxv}):
\begin{align*}
\dot{S}_{X|\mathcal{V}} &= S_{X|\mathcal{V}} B \Sigma_W B' S_{X|\mathcal{V}} - S_{X|\mathcal{V}} (A + B \Sigma_W B' S_X) \notag \\ &~~  - (A + B \Sigma_W B' 
S_X)'S_{X|\mathcal{V}} ,  %\label{eq:sxw}
\end{align*}
 if the respective initial condition $S_{0|\mathcal{V}}  = P_{0|\mathcal{V}}^{-1}$ is available as assumed in this proposition. 
\end{proof}
\end{prop}

Notice that the only difference from the IPF case is in the conditional covariance on the future verification entity; the modification from IPF amounts to calculation of the initial conditional covariance. Thus, once $P_{0|\mathcal{V}}$ is computed, the generalization in IWF does not incur additional complexity. 
However, calculation of this conditional covariance requires a sophisticated procedure, which is detailed in section \ref{sec:p0v}, in order to deal with noise-freeness of (fictitious) measurement of $V_t$ taken over finite time interval.

\begin{cor} \label{cor:ipf_equiv}
In case the verification variables are the whole state variables, i.e., $M_V = I_{n_X}$, the mutual information for (\ref{eq:iwf}) is reduced to that for (\ref{eq:ipf}):
$$
\mathcal{I}(\mathcal{X}_{[T_i, T_f]}; \mathcal{Z}_{[0,\tau]}) = \mathcal{I}(X_{T_i}; \mathcal{Z}_{[0,\tau]})
$$
where $ \mathcal{X}_{[T_i, T_f]} = \{ X_t : t \in [T_i, T_f]\}$. This can be shown as follows. The state history over $[T_i, T_f]$ can be decomposed as $\mathcal{X}_{[T_i, T_f]} = X_T \cup \mathcal{X}_{(T_i, T_f]}$. Notice that for any $X_t, t > T_i$, it is conditionally independent of  $\mathcal{Z}_{[0,\tau]}$ conditioned on $X_{T_i}$; thus, $\mathcal{I}(\mathcal{X}_{(T_i, T_f]}; \mathcal{Z}_{[0,\tau]}|X_{T_i}) =0$. Together with the chain rule of mutual information~\cite{inftheory}, this yields
\begin{align}
\mathcal{I}(X_{[T_i, T_f]}; \mathcal{Z}_{[0,\tau]}) &= \mathcal{I}(X_{T_i}; \mathcal{Z}_{[0,\tau]})  -  \mathcal{I}(\mathcal{X}_{(T_i, T_f]}; \mathcal{Z}_{[0,\tau]}|X_{T_i}) \notag \\& = \mathcal{I}(X_{T_i}; \mathcal{Z}_{[0,\tau]}). \notag
\end{align}
\end{cor}

\subsection{Calculation of Initial Conditioned Covariance $P_{0|\mathcal{V}}$} \label{sec:p0v}

The key difference between the mutual information for the windowed forecasting is that the initial conditional covariance $P_{0|\mathcal{V}}$ needs to be calculated conditioned on a random process rather than a finite-dimensional random vector. A typical way of calculating conditioned initial covariance is to pose a fixed-point smoothing problem with the fixed-point of interest being the initial time. But, note that for the windowed forecasting case, (\ref{eq:Vt}) plays a role of the measurement equation and thus there is no sensing noise in the measurement process. This lack of sensing noise prevents direct implementation of a conventional Kalman smoothing method (such as state augmentation) for computation of $P_{0|\mathcal{V}}$.

In this work, $P_{0|\mathcal{V}}$ is computed by the following three-step procedure:
\begin{enumerate}[leftmargin = .2in]
\item Calculate $P_{X}(T_i) = \Cov(X_{T_i} ) $ that is prior state covariance at $T_i$ by  integrating the Lyapunov equation:
\begin{equation}
 \dot{P}_X = A P_X + P_X A' + B \Sigma_W B' \label{eq:px}
\end{equation}
 over $[0, T_i]$ with  given initial condition $P_X(0) = P_0$.
\item Calculate $P_{X|\mathcal{V}}(T_i) \triangleq \Cov (X_{T_i} | \mathcal{V}_{[T_i, T_f]})$; this step is not trivial and section \ref{sec:pxvt} presents detailed derivation and procedure.
\item Compute $P_{0|\mathcal{V}}$ from $P_{X| \mathcal{V}} (T_i)$ by backward integrating the Lyapunov-like equation:
 \begin{align*}
 \dot{P}_{X|\mathcal{V}} = & (A + B \Sigma_W B'P_X^{-1}) P_{X|\mathcal{V}}  \\ &~~ + P_{X|\mathcal{V}} ( A + B\Sigma_W B'P_X^{-1})' - B \Sigma_W B'
\end{align*}
 from $T_i$ to $0$, coupled with (\ref{eq:px})~\footnote{The information form in (\ref{eq:sxv_w}) with (\ref{eq:sx_w}) can equivalently be used.}.
\end{enumerate}

\iffalse
\begin{rem}
Instead of computing $P_{0|\mathcal{V}}$, the conditional covariance at $\tau$ can also directly be computed from the procedure described in this section. This work explicitly computes $P_{0|\mathcal{V}}$ to allow for handling varying $\tau$.
\end{rem}
\fi

\subsubsection{Calculation of $P_{X | \mathcal{V}} (T_i)$ } \label{sec:pxvt}

Note that with perfect measurement of $V_t$ over some time interval, its time derivative can also be obtained. For example, $V_t = M_V X_t$ gives $\dot{V}_t = ( \dot{M}_V + M_V A ) X_t + M_V B W_t$; sensing of $V_t$ over $[T_i, T_f]$ is equivalent to sensing of $V_{T_i}$ and $\dot{V}_t$ over $[T_i, T_f]$. If $\dot{V}_t$ is again differentiable, i.e., $M_V B W_t = 0$, higher-order differentiation can also be reconstructed by repeatedly taking derivatives until the derivative contains white noise component. This observation was first identified by \cite{Bryson_TAC65} in the context of linear filtering with colored measurement noise and extended to a linear smoothing case~\cite{MehraBryson_TAC68}.

Suppose that $V_t$ is $K$-times differentiable, i.e., $K$-th derivative of $V_t$ contains white noise while lower-order derivatives do not contain white noise. Then,
\begin{eqnarray*}
V_t^{(k)} &=& H_{k}(t) X_t, \qquad k = 0, 1, \dots, K -1 \\
V_t^{(K)} & = & H_{K}(t) X_t + H_{K-1}(t) B W_t
\end{eqnarray*}
where $V_t^{(k)} \triangleq \frac{d^k V_t}{d t^k}$ and
\begin{eqnarray*}
H_{0}(t) = M_V(t), \qquad
H_{k+1}(t) =\dot{H}_{k}(t) + H_{k}(t) A(t).
\end{eqnarray*}
Note in this case that sensing of $\mathcal{V}_{[T_i, T_f]}$ is equivalent to sensing of $\mathcal{V}^{(K)}_{[T_i, T_f]} \bigcup \left( \bigcup_{i=0}^{K-1} V_{T_i}^{(k)} \right)$.
\begin{assum}
%It is assumed that differentiability of $V_t^{(i)}(t)$ is consistent throughout the verification time interval $[T_i, T_f]$. In other words,
The verification variables are assumed to satisfy:
\begin{enumerate}
\item ${\tt rank}( H_{k}(t) B(t) W_{t} )$ is either zero or full rank. Specifically, the rank is zero for $k < K$ and is $n_W$ for $k=K$.
\item ${\tt rank} ( H_{k}(t) B(t) W_{t} ) $ is time-invariant over $[T_i, T_f]$.
\end{enumerate}
\end{assum}
The first assumption can be easily relaxed by decomposing the verification variables into multiple sets depending on the order of differentiability. The second assumption readily holds for time-invariant $M_V$. For time-varying $M_V$, it does not hold in general; but, in this case the verification window can be decomposed into a sequence  of multiple sub-windows.

\begin{prop} \label{prop:smoothing}
The conditional covariance $P_{X|\mathcal{V}}(T_i) \triangleq \Cov(X_{T_i} | \mathcal{V}_{[T_i, T_f]})$ is given by:
\begin{equation}
P_{X|\mathcal{V}}(T_i) = \bar{P}(T_i) - \bar{P}(T_i) \Lambda(T_i) \bar{P}(T_i) \label{eq:psmooth}
\end{equation}
where $\bar{P}(t)$ and $\Lambda(t)$ are obtained by a system of matrix differential equations:
\begin{eqnarray}
\dot{\bar{P}} &=& \bar{A} \bar{P} +
\bar{P} \bar{A}' + B \bar{Q} B' - \bar{P} H_K \bar{R}^{-1} H_K' \bar{P} \label{eq:dot_pbar}\\
\dot{\Lambda} &= &- ( \bar{A} - \bar{P} H_K' \bar{R}^{-1} H_K )' \Lambda - \Lambda (\bar{A} - \bar{P} H_K' \bar{R}^{-1} H_K) \notag \\
&&~~~- H_K'  \bar{R}^{-1} H_K. \label{eq:dot_lambda}
\end{eqnarray}
where
\begin{eqnarray*}
\bar{R} & = & H_{k-1} B \Sigma_W B' H_{K-1}'\\
\bar{A} & = & A - B \Sigma_W B' H_{K-1}' \bar{R}^{-1} H_K\\
\bar{Q} & = & \Sigma_W - \Sigma_W B' H_{K-1}^T \bar{R}^{-1} H_{K-1} B \Sigma_W,
\end{eqnarray*}
and the boundary conditions are given as:
\begin{align*}
&\bar{P}(T_i) = P_X(T_i)  \notag \\ & \qquad  \qquad- P_X(T_i) \mathbf{H}' (T_i) \left[ \mathbf{H} P_{X}  \mathbf{H}'\right](T_i)^{-1} \mathbf{H} (T_i)  P_X(T_i), \\  %\label{eq:pbar_init} \notag\\
&\Lambda(T_f)  =  0, %\label{eq:lambda_tf} \notag
\end{align*}
where
$\mathbf{H} = [ H_0, \dots, H_{K-1}]'$.
\begin{proof}
See Appendix \ref{sec:proof_smoothing}
\end{proof}
\end{prop}

\begin{rem}
The system of matrix differential equations in (\ref{eq:dot_pbar}) and (\ref{eq:dot_lambda})  is a two-point boundary value problem for which boundary conditions for $\bar{P}$ are given at the initial time, $T_i$, and those for $\Lambda$ are given at the final time, $T_f$. However, since the Riccati equation for $\bar{P}$ is decoupled from the $\dot{\Lambda}$ equation in (\ref{eq:dot_lambda}), the system of equations can be solved in two steps:
\begin{enumerate}
\item The Riccati equation in (\ref{eq:dot_pbar}) is integrated forward over $[T_i, T_f]$ to obtain $\bar{P}(T_f)$,
\item The system of equations (\ref{eq:dot_pbar}) and (\ref{eq:dot_lambda}) are integrated together backwards to obtain $\Lambda(T_i)$.
\end{enumerate}
\end{rem}

\subsection{On-the-fly Information and Mutual Information Rate}
One benefit of exploiting conditional independence in computing mutual information is that it facilitates access to on-the-fly knowledge of how much information has been gathered and in what rate  information is being gathered.  Extending the development in \cite{ChoiHow_Auto10}  for IPF, on-the-fly information quantities can be obtained for IWF. For arbitrary time $t < \tau$, the mutual information accumulated up to $t$ is computed by:
$$
\mathcal{I}(\mathcal{V}_{[T_i, T_f]}; \mathcal{Z}_{[0,t]}) = \mathcal{J}_0^w(t) - \textstyle{\frac{1}{2}} \, \ldet
(I+Q_X(t) \Delta_S^w(t)).
$$
%\begin{rem}
Also, the rate of change of this on-the-fly information is obtained as
\begin{equation}
\textstyle{\frac{d}{dt}} \mathcal{I}(V_{[T_i, T_f]}; \mathcal{Z}_{[0,t]}) = \textstyle{\frac{1}{2}} {\tt tr} \left\{ \Sigma_N^{-1} C(t) \Pi^w(t) C(t)' \right\} \label{eq:iwf_rate}
\end{equation}
where $\Pi(t) = Q_X(t) \Delta_S^w(t)  \left[ I + Q_X(t) \Delta_S^w(t)  \right]^{-1} Q_X(t)$. The derivation is straightforward from the proof of mutual information rate for informative point-wise forecasting given in \cite[Proposition 4]{ChoiHow_Auto10}.
%\end{rem}
The mutual information rate in (\ref{eq:iwf_rate}) is particularly useful in order to visualize information distribution over space, because for many cases the observation matrix $C(t)$ is a function of the sensor location and thus the mutual information rate represents how high rate information can be obtained if sensing a certain location.

\section{Numerical Examples}
\subsection{Idealized Weather Forecasting Along Moving Paths}
The first example addresses design of continuous sensing trajectory for weather forecast improvement, when the simplified weather dynamic is described as the two-dimensional Lorenz-2003 model
\cite{Lorenz_JAS05, ChoiHow_GNC07}. The system equations of are 
\begin{align}
\dot{\phi}_{ij}  &= - \phi_{ij} - \zeta_{i-4,j} \zeta_{i-2,j} +
\textstyle { \frac {1} {3} \sum_{k \in [-1,1]} \zeta_{i-2+k,j}
\phi_{i+2+k,j} } \notag
\\ & - \mu \eta_{i,j-4} \eta_{i,j-2} + \textstyle {\frac
{\mu}{3} \sum_{k \in [-1,1]} \eta_{i,j-2+k} \phi_{i,j+2+k} } +
\phi_0 \notag
\end{align}
where $ \zeta_{ij} \triangleq \textstyle{ \frac {1} {3} \sum_{k \in [-1,1]}
\phi_{i+k,j} }, ~ \eta_{ij} \triangleq \textstyle{ \frac {1} {3} \sum_{k \in [-1,1]}
\phi_{i,j+k} } $ for $(i,j) \in \{1,2,\cdots, L_i\} \times \{1,2,\cdots, L_j \}$. The
subscript $i$ denotes the west-to-eastern grid index, while $j$ denotes the
south-to-north grid index; $\phi_{ij}$ represents some scalar meteorological variable at $(i,j)$-th grid. The boundary conditions of $\phi_{i+L_{i},j} =
\phi_{i-L_{i},j} = \phi_{i,j}$ and $\phi_{i,0} = \phi_{i,-1 } = 3$, $\phi_{i,L_{j}+1} =
0$ in advection terms, are applied to model the mid-latitude area of the northern
hemisphere as an annulus. The parameter values are $L_i=72, L_j=17, \mu =
0.66$ and $\phi_0=8$. The size of $1 \times 1$ grid corresponds to 347 km
$\times$ 347 km in real distance, and unit time in this model is equivalent to 5
days in real time. The overall system is tracked by a nonlinear estimation
scheme, specifically an ensemble square-root filter (EnSRF)~\cite{ensrf} data
assimilation scheme, that incorporates measurements from a fixed observation
network of size 186.

The path planning problem is posed for a linearized model over
some $4 \times 3$ local region (therefore, $n_X = 12$) in the
entire $L_i \times L_j$ grid space. A linear time-invariant model is
obtained by deriving the Jacobian matrix of the dynamics around the
nonlinear estimate for $\phi_{ij}$'s at the grid points in the local
region. Thus, the state vector $X_t$ represents the perturbation of
$\phi_{ij}$'s from the ensemble mean:
$$
X_t = \left[ \delta \phi (\mathbf{r}_1), \dots, \delta \phi (\mathbf{r}_{n_X}) \right]'
$$
where $\mathbf{r}_k,~k=1,\dots, n_X$ represents the location vector of $k$-th grid point, and $\delta \phi_t$ denotes the perturbation of the Lorenz variable.  The prior covariance  of the state, $P_0$ is provided EnSRF data assimilation scheme.

%In this linear model, the dependence of the local dynamics on the evolution of the external
%dynamics is ignored in deriving the Jacobian matrix (or $A$ matrix).
%Instead, this effect is incorporated in the process noise term,
%i.e., the states on the boundary of the local region, which may be
%affected by external dynamics more substantially, are assumed to be
%subject to larger process noise.

The continuous trajectory planning allows for sensing at an off-gird point. To represent measurement at an off-grid point, squared-exponential kernel function often used in Kriging\cite{krig} and Gaussian process regression is adopted; observation variable at some location $\mathbf{r}$ is expressed as a linear combination of the state:
\begin{align}
\delta \phi_t(\mathbf{r}) %&= \sum_{i=1}^{n_X}  \sum_{j=1}^{n_X} \alpha_{ij} \rho (\mathbf{r}, \mathbf{r}_j ) \delta \phi (\mathbf{r}_i) \notag \\
 %&
  = \left[ C_{1}~\dots ~ C_{i} ~\dots~ C_{n_X}\right] X_t \label{eq:phi_t}
\end{align}
with $ C_{i} = \textstyle{\sum_{j=1}^{n_X}}
\alpha_{ij} \rho(\mathbf{r},\mathbf{r}_j)$, where
 $ \rho(\mathbf{r},\mathbf{r}_j) = \exp\big\{
 -(x-x_j)^2/(2 l_x^2) -(y-y_j)^2/(2 l_y^2) \big\}
$ and $\alpha_{ij}$ is the $(i,j)$-th element of the matrix $\left[ \rho(\mathbf{r}_i, \mathbf{r}_j ) \right]^{-1} $.  The length-scale of $(l_x,l_y) = (1,0.7)$ is used in this example.

The goal of planning is to design a 6-hr flight path ($\tau=$ 6 hrs) for a single UAV sensor platform to improve
forecast over two moving verification object of interest between $T_i = $ 60hrs and $T_f=$ 84hrs; the two objects are to move one grid ($\sim 347$km) over this verification window.
The motion of the sensor platform is described as 2-dimensional holonomic motion:
\be
\dot{x} = v \cos \theta, \qquad \dot{y} = v \sin \theta \label{eq:motion}
\ee
with constant speed $v =1/3$ grid/hr ($\sim 116$km/hr); the path angle $\theta$ is the control variable to optimize on; along the trajectory the sensor continuously measure $\phi_t(\mathbf{r})$ in (\ref{eq:phi_t}) with additive Gaussian noise with $\Sigma_N = 0.05^2$.

The verification path is assumed to be a straight line starting at some point and ending at another point, each of which corresponds to some linear combination of the state variables.
$$
V_t = \left[ \frac{T_f - t}{T_f - T_i} M_V^i + \frac{t - T_i}{T_f - T_i} M_V^f \right] X_t
$$
where $M_V^i $ and $M_V^f$ are the verification matrices for $T_i$ and $T_f$, respectively. This allows for computing the time derivative of verification matrix:
$$
\dot{M}_V =( - M_V^i + M_V^f)/(T_f - T_i),
$$
which will be needed to construct $H_1$ matrix in section \ref{sec:p0v}.

Four different planning strategies are compared: (a) optimal IWF trajectory, (b) gradient-ascent steering with IWF information potential field, (c) optimal trajectory for IPF problem with $T = (T_i + T _f)/2$ and $M_V = (M_V^i + M_V^f)/2$, and (d) gradient-ascent steering associated with IPF mutual information. To obtain the optimal solution, the control is parameterized as a piece-wise linear function of time with 6 segments and the 36 straightline paths with different flight path angles are considered as initial guess to the optimization; for solving resulting nonlinear programs, TOMLAB/SNOPT~\cite{snopt} is used. 
For the two-dimensional holonomic sensor motion in
(\ref{eq:motion}), the gradient-ascent steering law is obtained by taking the partial derivative of the information rate (in \ref{eq:iwf_rate} for IWF), which is a function of spatially-dependent $C$ matrix, with respect to the spatial coordinates:
\be \textstyle %
\theta_{G}(t) = \tan^{-1} \left\{ \frac{ \Sigma_N^{-1} C(x(t), y(t)) \Pi^w(t) \mathbf{d}(y)}{\Sigma_N^{-1} C(x(t), y(t)) \Pi^w(t) \mathbf{d}(x)} \right\}
\notag
\ee %
where $\mathbf{d}(x)$ is an $n_X$-dimensional column vector with $\mathbf{d}(x)_i = - l_x^{-2} \sum_j \alpha_{ij} \rho(\mathbf{r},\mathbf{r}_j) (x - x_j)$ (and $\mathbf{d}(y)$ is defined similarly.)

\begin{figure}[t]
\centerline{\includegraphics[width=1\columnwidth, trim=43 30 45
23,clip]{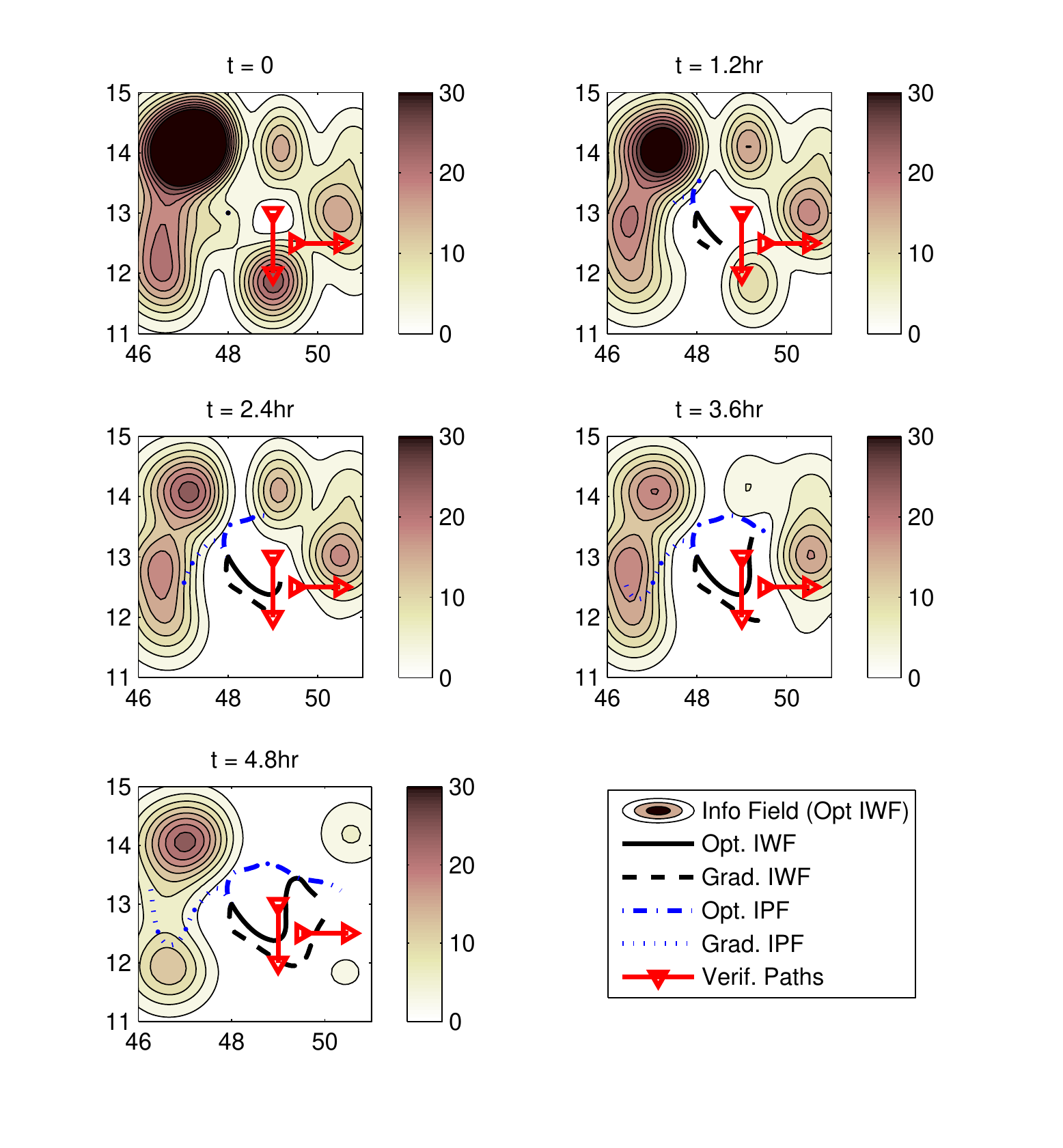}}
\vspace*{-.15in}
  \caption{Trajectories by different strategies overlaid with information potential field for the optimal IWF trajectory}
  \label{fig:snapshot}
 % \vspace*{-.02in}
\iftrue
\centerline{\includegraphics[width=1\columnwidth, trim=20 147 30 15,clip]{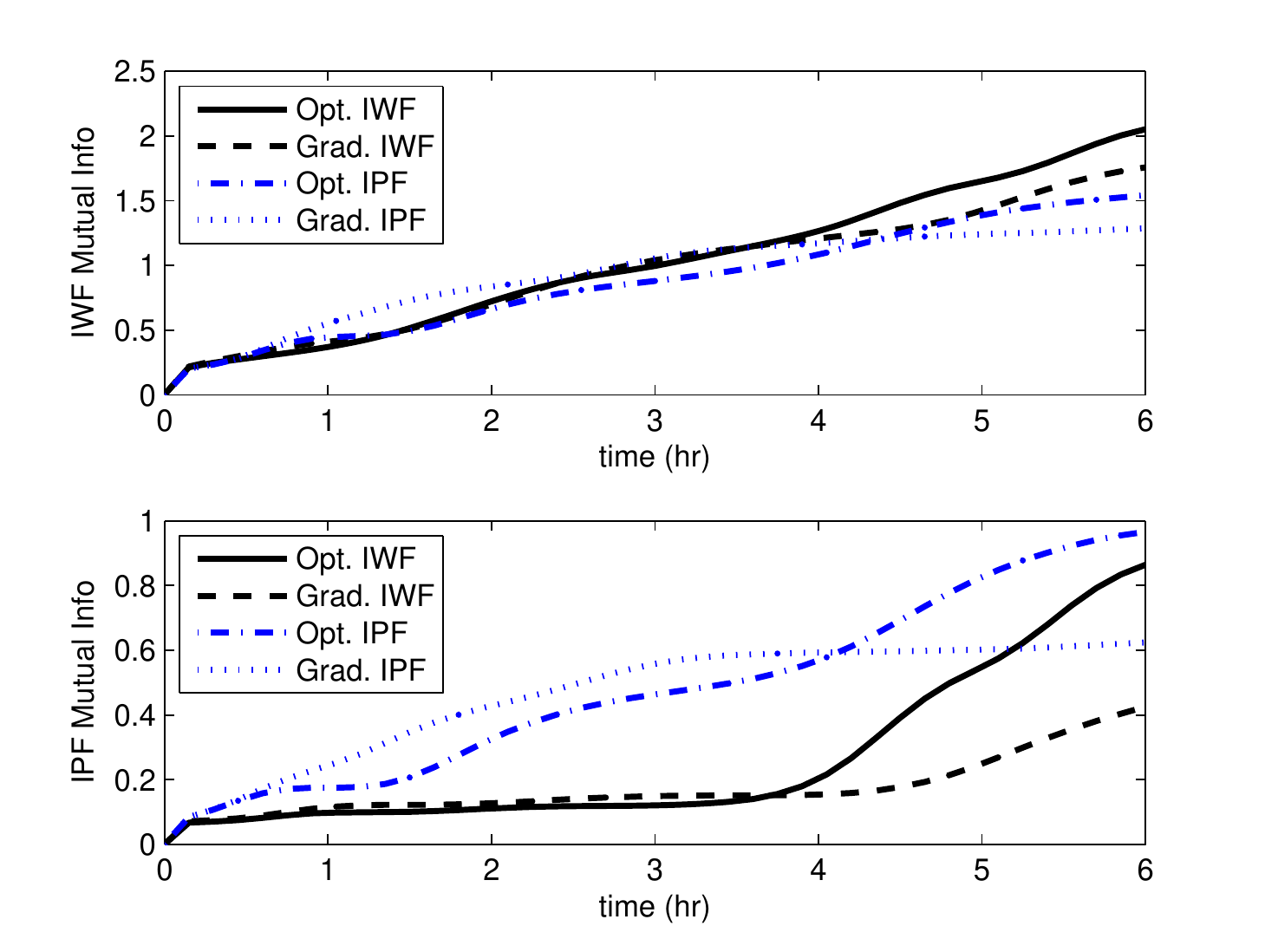}}
\vspace*{-.15in}
  \caption{Information accumulation along different sensing trajectories}
  \label{fig:info_history}
\fi
  \vspace*{-.0in}
\end{figure}

Fig. \ref{fig:snapshot} illustrates snapshots of trajectories of the four planning strategies overlaid on the information potential field associated with the optimal IWF solution. Two verification paths are depicted with red triangles while the shape of triangle is aligned with the direction of movement of the verification paths. In the potential field, a dark part is more informative than brighter one; note that his potential field differs from that of the IWF gradient ascent solution except for the initial time, as $\Pi^w(t)$ in (\ref{eq:iwf_rate}) depends on the sensing choice up to $t$. Although not depicted, in this particular example, the IPF potential field looks substantially different from the IWF field in that the lower information peak around $(49, 12)$ does not exist for the IPF case. That may be the reason the IPF-based trajectories move left up while the IWF-based trajectories start moving right down. Fig. \ref{fig:info_history} illustrates information accumulation along the four trajectories. Notice that the optimal IWF solution might not provide the most information earlier in the trajectory, but results in gathering most amount of information in the end.

\iffalse
\begin{table}[t]
\begin{center} %\setlength{\extrarowheight}{4pt}
 \caption{ Mutual information values for different strategies } \label{tab:mutual_info_values}
\begin{tabular}{c||c|c|c |c }
Strategies & IWF Opt. & IWF Grad. & IPF Opt. & IPF Grad. \\ \hline \hline
IWF Info & 0.8642 &0.4252 &  0.9657& 0.6238\\
IPF Info &2.0511  &  1.7567&1.5409    &1.2853
\end{tabular}
\end{center}
%\vspace*{-.15in}
\end{table}
\fi

\subsection{Sensor Network Management for Target Tracking}
The second example considers management of fixed sensor networks deployed in a two-dimensional space for tracking moving targets. There are $n_T$ targets and each target's motion is represented by the Singer model~\cite{Li_TAES03}, i.e., kinematic model with first-order Markov diffusion in acceleration:
\begin{equation}
\begin{split}
&\dot{x}^k = v_x^k,~~\dot{v}_x^k = a_x^k, ~~\dot{a}_x^k = - \kappa a_x^k + w_x^k \\
&\dot{y}^k = v_y^k,~~\dot{v}_y^k = a_y^k, ~~\dot{a}_y^k = - \kappa a_y^k + w_y^k
\end{split} \label{eq:singer}
\end{equation}
where the superscript $k$ denotes $k$-th target and $\kappa = 0.4$ is used in this work. The state vector is defined as $X_t = \left[ [x^k~v_x^k~a_x^k~y^k~v_y^k~a_y^k]', ~k = 1, \dots, n_T \right]$ and its initial covariance $P_0$ is computed from an extended Kalman filter run with randomly chosen measurements.

The sensor network consists of $n_S$ sensors that can measure the pseudo-range between themselves and the targets. The linearized pseudo-range measurement sensing model is given by $Z_t^s = [Z_{t}^{s,k}, ~k=1,\dots, n_T]'$ with
$$
Z_{t}^{s,k} = - \frac{2 \alpha}{|| \bar{\mathbf{r}}_k - \mathbf{l}_s ||^2 + \beta } (\mathbf{l}_s - \bar{\mathbf{r}}_k )' \left[ x^k ~~ y^k \right]'
$$
where $\mathbf{l}_s$ is the location vector of $s$-th sensor, $\bar{\mathbf{r}}_k$ is the nominal position of the $k$-th target; this linearized model is obtained by calculating the Jacobian of the pseudo-range, $ \alpha/ (|| \mathbf{r}_k - \mathbf{l}_s ||^2 + \beta )$ around $\bar{\mathbf{r}}_k$; $\alpha = 2000, ~\beta = 100$ is used in this example.

The sensor management problem is to determine which $m_s$ sensors (out of $n_S$) to turn on to best track the targets; the total planning horizon $[0,\tau]$ is dividend into $m_\tau$ intervals within which the same set of sensors are turned on to make continuous observation of the targets. To goal of this management is to reduce uncertainty in  \textit{velocities} of the targets over the verification window. Problem parameters in this example are as follows: $n_S = 20$, $m_s = 5$, $\tau = 5$, $m_\tau = 5$, $T_i = 3$, $T_f = 5$, $\Sigma_W = 0.07^2 $, and $\Sigma_N = 0.25^2 $. The verification matrix in this case is given by $M_V = \mathbf{1}_{n_T}' \otimes [0~1~0~0~1~0]$ where $\otimes$ denotes the Kronecker product.

With the system described in (\ref{eq:singer}), the verification variables are twice differentiable, in other words, perfect measurement of the velocity over some finite window is equivalent to perfect sensing of velocity and acceleration at $T_i$ and noisy sensing of the jerk. For this reason, the initial value of the Riccati matrix for the smoothing, i.e., $\bar{P}(T_i)$ has non-zero elements in columns and rows corresponding only to the position variables.

\begin{rem}
Note that the position dynamics does not contain white noise, so does the velocity dynamics. Therefore, by taking perfect measurement of position over $[T_i, T_f]$, the velocity and the acceleration can also be determined perfectly; in other words, if with perfect position information over some finite time window, the entire state can be reconstructed. Therefore, although the problem is formulated as an informative windowed forecasting, this is equivalent to the case in Corollary \ref{cor:ipf_equiv}: $\mathcal{I}(\mathcal{V}_{[T_i, T_f]}; \mathcal{Z}_{[0,\tau]}) = \mathcal{I}(X_{T_i}; \mathcal{Z}_{[0,\tau]})$.
\end{rem}

\begin{figure}[t]
\centerline{\includegraphics[width=1\columnwidth, trim=60 20 0
15,clip]{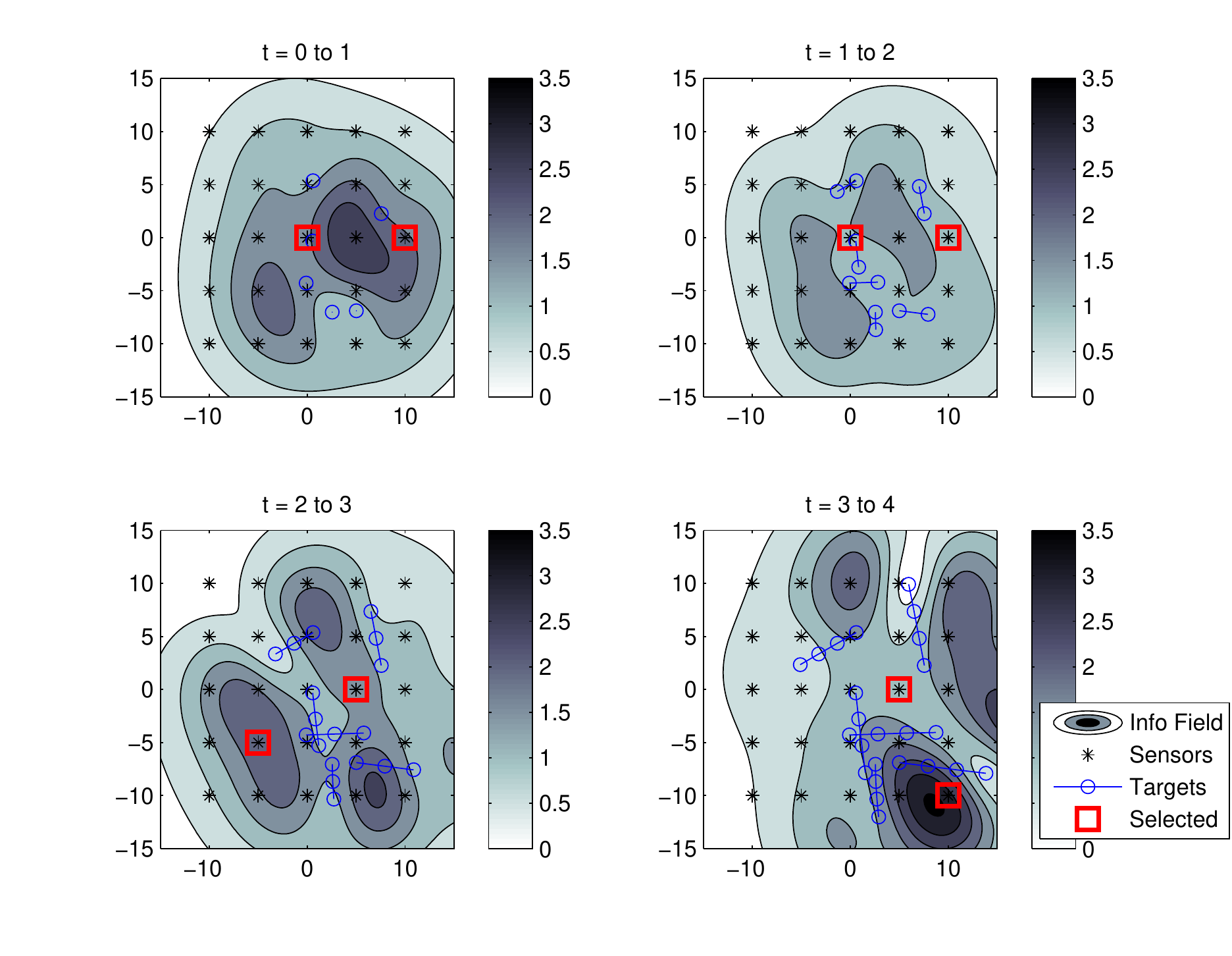}}
\vspace*{-.15in}
  \caption{Optimal sensor schedules overlaid on information field}
  \label{fig:sn_tracks}
  \vspace*{-.05in}
\end{figure}

Fig. \ref{fig:sn_tracks} illustrates the optimal scheduling of sensor networks overlaid with the information potential calculated at the start time of the respective time period. It can be found that the optimal selection for each period does not necessarily selects the point with the highes potential point as the selection is made for the best combination of two sensors that is most informative for predicting the future. Table \ref{tab:snet_accum} summarizes the mutual information value acquired during each decision period; the same quantities for the cases when target positions or accelerations are the verification entity are also shown for comparison. In this particular example, the selected sensor sequences are not significantly different for different choices of verification entity (in particular for the first two decision periods). 

\iftrue
\begin{table}[t]
\begin{center} %\setlength{\extrarowheight}{4pt}
 \caption{Information gains per decision period} \label{tab:snet_accum}
\begin{tabular}{c||c|c|c |c }
Case  & $t=$ 0 to 1  & $t=$ 1 to 2  & $t=$ 2 to 3  &  $t=$ 3 to 4 \\ \hline \hline
Pos. &       6.9507  &     6.1875  &      6.0211 &    5.9962 \\
Vel. &     2.6486    &      2.7414 &      2.6148  &     2.4116  \\
Acc. &      0.0384 &     0.0531 &     0.0527 &     0.0335
\end{tabular}
\end{center}
\vspace*{-.1in}
\end{table}
\fi

Observe that the mutual information value is greatest for the position case and smallest for the acceleration case. This can be well explained on the basis of commutativity of mutual information: $\mathcal{I}(\mathcal{V}_{[T_i,T_f]}; \mathcal{Z}_{[0,\tau]}) =\mathcal{I}(\mathcal{Z}_{[0,\tau]}; \mathcal{V}_{[T_i,T_f]})$. Namely, informativeness of measurement to predict verification entity is equivalent to informativeness of the verification entity to predict the measurement. With the kinematic model in this example, knowing future position over finite window means knowing every kinematic variables (likewise, knowing velocity means knowing acceleration as well); thus, position is more informative than velocity, and acceleration is nearly non-informative as it is corrupted by white noise.

\section{Conclusions}\label{sec:conclusions}
This paper has presented a formula of mutual information in the informative forecasting problem that is to reduce uncertainty in the future verification variables over finite time window. While underlying conditional independence relation has allowed for exploiting the smoother form of mutual information, fixed-interval smoothing with perfect measurement is utilized to compute the conditional initial covariance matrix that provides initial conditions for required matrix differential equations. Numerical examples highlighting temporal variation  and differentiability of the verification entity have demonstrated the validity of the proposed approach.

%%%%%%%%%%%%%%%%%%%%%%%%%%%%%%%%%%%%%%%%%%%%%%%%%%%%%%%%%%%%%%%%%%%%%%%%%%%%%%%%
\begin{ack}
This work was supported in part by AFOSR grant (FA9550-12-1-0313), and in part by the KI Project via KI for Design of Complex Systems.
\end{ack}

%%%%%%%%%%%%%%%%%%%%%%%%%%%%%%%%%%%%%%%%%%%%%%%%%%%%%%%%%%%%%%%%%%%%%%%%%%%%%%%%

\appendix
\section{Proof of Proposition \ref{prop:smoothing}} \label{sec:proof_smoothing}
This section proves Proposition \ref{prop:smoothing} by solving an optimal control problem equivalent to the fixed-interval smoothing problem to compute $\Cov(X_{T_i}|\mathcal{V}_{[T_i, T_f]})$. While the key concept in the procedure is adopted from \cite{MehraBryson_TAC68, BrysonHo75}, extended expression for high-order differentiable $V_t$ is newly detailed herein. 

As \cite{MehraBryson_TAC68, BrysonHo75} pointed out, a fixed-interval smoothing problem can be viewed as an optimal control problem to find out optimal (continuous) sequence of process noise over the interval and the optimal initial state under the constraint of measurements obtained over the interval. Therefore, the smoothing problem in section \ref{sec:p0v}  can equivalently be written as an optimal control problem with associated adjoint variables:
\begin{equation}
\begin{split}
J = &\frac{1}{2} X_{T_i}' P_{T_i}^{-1} X_{T_i} +  \sum_{k=0}^{K-1} \nu_k'  ( V_{T_i}^{(k)} - H_{k}(T_i) X_{T_i}) \\
& + \int_{T_i}^{T_f} \left\{ \frac{1}{2} W_t' \Sigma_W^{-1} W_t  + \lambda(t)' \left( \dot{X}_t - A  X_t - B  W_t \right) \right. \\
& + \left. \mu(t)' \left( V_t^{(K)} - H_{K}  X_t - H_{K-1} B W_t \right) \right\} dt
\end{split} \notag
\end{equation}
Although not explicitly indicated for brevity, all the system matrices can be time-varying. In this optimal control problem, the process noise sequence $W_t$ plays a role of control input. The adjoint variables $\lambda(t) \in \mathbb{R}^{n_X}$, $\mu(t) \in \mathbb{R}^{n_V}$, and $\nu_i \in \mathbb{R}^{n_V}$ correspond to the dynamics, measurement of white-noise-containing derivative of $V_t$ throughout the verification interval, and measurement of white-noise-free derivative of $V_t$ at the initial time.

The optimality condition can be written as
\begin{eqnarray*}
W_t^{\star}  &= & \Sigma_W B' \left( \lambda + H_{K-1}' \mu \right) \label{eq:Wtopt}\\
\dot{\lambda} &=& - A' \lambda - H_{K}' \mu \label{eq:lambda_dot} \\
\dot{X}_t &=& A X_t + B W_t^{\star} \label{eq:XdoT_ipt}
\end{eqnarray*}
with the boundary conditions
\begin{eqnarray}
\lambda(T_i) &=& P_{T_i}^{-1} X_{T_i} - \textstyle \sum_{k=0}^{K-1} H_{k}' \nu_i  \label{eq:init_lambda}\\
\lambda(T_f) & = & 0.  \label{eq:final_lambda}
\end{eqnarray}

The adjoint vector $\mu(t)$ can be expressed as a function of the state and the costate from the white-noise-containing measurement equation.
$$
V_t^{(K)} - H_{K} X_t - H_{K-1} B \Sigma_W B' \left( \lambda + H_{K-1}' \mu \right) = 0,
$$
which leads to
\be %
\mu = \bar{R}^{-1} \left( - H_K X_t - H_{K-1} B \Sigma_W B' \lambda + V_t^{(K)} \right). \label{eq:mu_opt}
\ee %
Plugging (\ref{eq:mu_opt}) back into the state and the costate dynamic results in:
\begin{equation}
\begin{bmatrix} \dot{X}_t \\ \dot{\lambda}_t \end{bmatrix}
= \begin{bmatrix} \bar{A} & B \bar{Q} B' \\
H_K' \bar{R}^{-1} H_K  & - \bar{A}'
\end{bmatrix} \begin{bmatrix} X_t \\ \lambda_t \end{bmatrix}
+ \begin{bmatrix} \bar{G}  \\ - H_K' \bar{R}^{-1}  \end{bmatrix} \dot{V}^{(K)} \notag \label{eq:hamiltonian}
\end{equation}
where
\begin{eqnarray*}
\bar{R} & = & H_{K-1} B \Sigma_W B' H_{K-1}' \\
\bar{F} & = & A - B \Sigma_W B' H_{K-1}' \bar{R}^{-1} H_K\\
\bar{Q} & = & \Sigma_W - \Sigma_W B' H_{K-1}' \bar{R}^{-1} H_{K-1} B \Sigma_W \\
\bar{G} & = & B \Sigma B' H_{K-1}' \bar{R}^{-1}
\end{eqnarray*}
%The boundary condition for this system of differential equations is given only for the costate variables. 
The initial condition in (\ref{eq:init_lambda}) can be rewritten as:
$$
X_{T_i} = P_{T_i} \left( \lambda_{T_i} + \textstyle \sum_{k=0}^{K-1} H_i(T_i)' \nu_i \right).
$$
Note that from the measurement equations at $T_i$, the following system of equations can be obtained:
$$
V^{(K)}_{T_i} = H_k(T_i) P_{T_i} \lambda_{T_i} + H_k (T_i) P_{T_i}\textstyle \sum_{j=0}^{K-1} H_j(T_i)' \nu_j
$$
for $k = 0, \dots, K-1$; this leads to
$$
\boldsymbol{\nu} = \left[ \mathbf{H} P \mathbf{H}'\right](T_i)^{-1} \left[  \mathbf{dV}_{T_i} - \mathbf{H}(T_i) P_{T_i} \lambda_{T_i}  \right]
$$
where $ \mathbf{dV} = [ V, \dot{V}, \dots, V_t^{(K)} ]'$, $\mathbf{H} = [ H_0, \dots, H_{K-1}]'$. Therefore, the optimal estimate of initial state is given by
\begin{equation}
\begin{split}
X_{T_i} = & \underbrace{P_{T_i} \mathbf{H}(T_i)' \left[ \mathbf{H} P \mathbf{H}'\right](T_i)^{-1}\mathbf{d V}_{T_i}}_{\widehat{X}_{T_i}^{+}} \\
& + \underbrace{\left[ P_{T_i} - P_{T_i} \mathbf{H}(T_i)' \left[ \mathbf{H} P \mathbf{H}'\right](T_i)^{-1} \mathbf{H} P_{T_i} \right]}_{P_{T_i}^{+}} \lambda_{T_i}.
\end{split} \notag
\end{equation}
As indicated in \cite{MehraBryson_TAC68}, $P_{T_i}^{+}$ is corresponds to the conditional covariance of $X_{T_i}$ conditioned on the perfect measurement of white-noise-free derivatives of $V_t$ at $T_i$.  With this setting, the smoothed estimate of the state, i.e., $\widehat{X}_{t} \triangleq \mathbb{E}[X_t | \mathcal{V}_{[T_i, T_f]}]$ can be obtained by
\begin{equation}
\widehat{X}_{t}^S = \widehat{X}_t^F + P_F(t) \lambda_t \notag
\end{equation}
where $\widehat{X}_t^F = \mathbb{E}[X_t | \mathcal{V}_{[T_i, t]} ]$ is the filtered stated estimate and $P_F$ is the error covariance of the filtered state estimated.
The filtered covariance, i.e., $\Cov(X_t | \mathcal{V}_{[T_i, t]} )$ can be obtained from the Riccati equation:
$$
\dot{P}_{F} =
\bar{A} P_{F} +
P_{F} \bar{A}' + B \bar{Q} B' - P_{F} H_K \bar{R}^{-1} H_K' P_F
$$
with initial condition $P_F(T_i) = P_{T_i}$. Although not needed for the context of windowed forecasting where the actual measurement values are not available at the decision time, the filtered estimate is obtained by
$$
\widehat{X}_t^F = \bar{A} \widehat{X}_t^F + P_F H_K' \bar{R}^{-1} \left( V_t^{(K)} - H_K \widehat{X}^F_t \right) + \bar{G} V_t^{(K)}
$$

The covariance of the smoothed estimate can be obtained by
$$
P_{X|\mathcal{V}} = P_F - P_F \Lambda P_F
$$
where
\begin{equation}
\begin{split}
\dot{\Lambda} = &- ( \bar{A} - P_F H_K' \bar{R}^{-1} H_K )' \Lambda - \Lambda (\bar{A} - P_F H_K' \bar{R}^{-1} H_K) \\
&~~- H_K' \bar{R}^{-1} H_K
\end{split} \notag
\end{equation}
with the terminal condition $\Lambda(T) = 0$.

\renewcommand*{\bibfont}{\footnotesize}

\bibliographystyle{elsart-harv}
\bibliography{iwf_automatica}

\end{document}